\documentclass[a4paper,pra,onecolumn,showpacs,superscriptaddress,groupedaddress]{revtex4}
\usepackage{ae}
\usepackage[T1]{fontenc}
\usepackage[ansinew]{inputenc}
\usepackage{amsmath}
\usepackage{amsbsy}
\usepackage{amssymb}
\usepackage[caption=false]{subfig}
\usepackage{multirow}
\usepackage{array}
\usepackage[]{graphicx}
\usepackage{color}
\usepackage[colorlinks]{hyperref}
\usepackage{lscape}
\newtheorem{theorem}{Theorem}

\begin{document}
\title{Coherence Fraction}
\author{Sumana Karmakar}
\email{sumanakarmakar88@gmail.com}
\affiliation{Department of Applied Mathematics, University of Calcutta, 92, A.P.C. Road, Kolkata-700009, India.}
\author{Ajoy Sen}
\email{ajoy.sn@gmail.com}
\affiliation{Department of Applied Mathematics, University of Calcutta, 92, A.P.C. Road, Kolkata-700009, India.}
\author{Indrani Chattopadhyay}
\email{icappmath@caluniv.ac.in, ichattopadhyay@yahoo.co.in}
\affiliation{Department of Applied Mathematics, University of Calcutta, 92, A.P.C. Road, Kolkata-700009, India.}
\author{Amit Bhar}
\email{bhar.amit@yahoo.com}
\affiliation{Department of Mathematics, Jogesh Chandra Chaudhuri College, 30, Prince Anwar Shah Road, Kolkata-700033, India.}
\author{Debasis Sarkar}
\email{dsappmath@caluniv.ac.in, dsarkar1x@gmail.com}
\affiliation{Department of Applied Mathematics, University of Calcutta, 92, A.P.C. Road, Kolkata-700009, India.}
\begin{abstract}
The concept of entanglement fraction is generalized to define coherence fraction of a quantum state. Precisely, it quantifies the proximity of a quantum state to maximally coherent state and it can be used as a measure of coherence. Coherence fraction has a connection with $l_1$-norm coherence and provides the criteria of  coherence distillability. Optimal coherence fraction corresponding to a channel, defined from this new idea of coherence fraction, obeys a complementary relation with its decohering power. The connection between coherence fraction and $l_1$-norm coherence turns to hold for bipartite pure states and $X$ states too. The bipartite generalization shows that the local coherence fractions of a quantum state are not free and they are bounded by linear function of its global coherence fraction. Dynamics of optimal coherence fraction is also studied for single sided and both sided application of channels. Numerical results are provided in exploring properties of optimal coherence fraction.
\end{abstract}
\maketitle

\section{Introduction}
Entanglement has been established as one of the prominent resources in quantum theory, providing ways to perform tasks better than or sometimes impossible by classical means and thus demarcating the classical limits\cite{ent1,ent2}. One of the main ingredients behind the creation of entanglement is coherent superposition of quantum states. The idea of coherence dates back to classical optics as a property of wave. However, in quantum domain the idea has been elevated to a governing principle. Since superposition depends on the measured basis, coherence is defined with respect to some fixed basis. Coherence could be viewed as the manifestation of quantumness without invoking correlations that involves composite quantum systems like, entanglement, discord, etc. Quantum coherence is the resource behind several quantum mechanical phenomenon. The role of coherence has been confirmed in biological systems\cite{bio1,bio2,bio3}, quantum thermodynamics \cite{thermo1,thermo2,thermo3,thermo4,thermo5}, nanoscale physics, and other related tasks \cite{smc1}.

In order to establish quantum coherence as a useful resource it is inevitable to develop a proper quantification scheme. Attempts have been taken to construct a suitable measure of quantum coherence based on different frameworks like skew information \cite{coh2}, entropic and distance based approaches \cite{coh1,coh3}. A resource theoretic view point of quantum coherence has also been introduced by Baumgratz \textit{et al.} \cite{coh1}. Some recent works in \cite{coh resource-win, coh resource-chit, coh distillation, assisted distillation,coh robustness} established a rigorous mathematical framework for resource theory  of coherence. Relationships between  quantum coherence and other resources of quantum information theory like entanglement\cite{coh-ent}, mixedness\cite{coh-mix}, discord, deficit\cite{coh-correlation} etc., have also been studied.

In quantum domain noises are modelled generally by quantum channels. Resources must be guarded against such noises in order to use them in resource theoretic tasks. Hence we need investigation on the behaviour of resources under different channels those induce noises.  Dynamical evolution of quantum coherence under different quantum channels is an important area of investigation. Recent researches established that quantum coherence can freeze \cite{freezing} under noisy environment. Characterization of quantum channels, according to their coherence manipulating ability, evolves along this way. It involves the study of channels which could preserve coherence \cite{preserving} and the channels that could not generate coherence from incoherent state \cite{non-generating}. We find cohering and decohering power\cite{coh pow1, coh pow2, coh pow3} of quantum channels which quantifies the ability of creating or destroying coherence from quantum states. Cohering power of a general quantum evolution has been studied too\cite{cohpowqe}.

In entanglement theory, entanglement fraction \cite{ent frac1,ent frac2,ent frac3,ent frac4,ent frac5,ent frac6} quantifies the entanglement content of a state by measuring its overlap with maximally entangled state. This quantity has a deep connection to various information theoretic processes like dense coding, teleportation, etc., \cite{ent frac7}. For example, a two-qubit quantum state is able to teleport qubit iff its entanglement fraction is greater than $\frac{1}{2}$. In this paper, we have developed the idea of coherence fraction of a quantum state in the theory of quantum coherence that quantifies the maximum overlap of a quantum state with a maximally coherent state \cite{preserving,MCS1}. Considering the task of quantifying coherence fraction of a state we study its properties and explain its resource theoretic connections. Whenever a quantum state is subjected to noises, it is bound to loose its coherence. Naturally the question arises that to what extent a quantum channel can affect the coherence of a state? We have tried to answer of this question in this work. We have also defined the maximum achievable coherence fraction of a quantum channel. These newly developed ideas can be used in characterizing quantum channels and will ultimately explain the effectiveness of a quantum channel in sustaining coherence. Some of the problems in entanglement theory have been studied in ref. \cite{ef}.

Our paper is organized as follows. In section II we will discuss some preliminary ideas of coherence. We will introduce the idea of coherence fraction in section III and establish some connections with resource theory of coherence. The idea of optimal coherence fraction introduced in Section IV and the idea will be explained by some examples of important channels. We will also discuss the complementary relation with decohering power. Section V deals with the extension of all the ideas to bipartite system, proving some new results and describing the dynamics through some examples. Finally, we will conclude in Section VI.

\section{Preliminaries and notations}
Like any other resource theory, resource theory of coherence is built up with three main ingredients- coherent states, incoherent states and incoherent operations. Since superposition depends on the particular basis, the idea of coherence needs to fix the basis first. Depending on the particular pre-chosen basis, any quantum state which is diagonal in the basis is called incoherent state as it has no power to produce superposition. Mathematically, all incoherent states have the form $\delta=\sum_{i=1}^d\delta_i|i\rangle\langle i|$ where $\delta_i\in[0,1]$ and $\sum_{i=1}^d\delta_i=1$. The set of all incoherent states is convex and it is denoted by $\mathcal{I}$. All other states are called coherent states. Now, by a quantum operation we mean a completely positive trace preserving (CPTP) map $\Phi$ which can be represented in Kraus form as  $\Phi(\rho)=\sum_i{K_i\rho K_i^{\dagger}}$  where $\sum_iK_i^{\dagger}K_i=\textit{I}$ and $\{K_i\}$'s are called the Kraus operators of the quantum operation $\Phi$. Incoherent operations are precisely the quantum operations, having incoherent Kraus operators, \textit{i.e.,} $K_i\mathcal{I}K_i^{\dagger}\subset\mathcal{I}$ for all $i$.  Each incoherent Kraus operators maps incoherent state to another incoherent state. Incoherent operations are not unique and they have other finer classifications (e.g., genuine incoherent, strict incoherent, etc., \cite{thermo2,coh1,ic3,ic4,ic5}). Usually, coherence has been quantified through distance-based quantifiers (like, $l_1$ norm, trace norm, Bures distance, etc.) and also by several other means, such as relative entropy, robustness, skew information, etc. The resource theory of coherence put forward two new coherence measures- distillable coherence and coherence cost, similar to the counterpart in entanglement theory.  In general, it is expected that any coherence measure should be non-negative, convex, monotonic and must vanish on incoherent states. In this paper, we will mainly consider $l_1$-norm coherence which is defined \cite{coh1} via the off diagonal elements of a density matrix $\rho$ in the reference basis (usual incoherent computational basis ($\{|i\rangle\}$) is used throughout the paper in defining coherence) as
\begin{equation}
C_{l_1}(\rho):=\sum_{i\ne j,i,j=0}^{d-1}|\langle i|\rho|j\rangle|.
\end{equation}
This measure satisfies all the necessary requirements of a bona fide coherence measure. Relative entropic coherence is another good measure of coherence, having resource theoretic interpretation. It is defined as \cite{coh1}
$$C_r(\rho):=S(\Delta(\rho))-S(\rho),$$
where $S$ is the von Neumann entropy and $\Delta(\rho)$ is the diagonal part of the state $\rho$. Intrinsic randomness corresponding to a measurement can be used to define coherence as well \cite{intrand}.  Intrinsic randomness ($R_{I}(|\psi\rangle)$ of projective measurement $P_{I}=\{P_i=|i\rangle\langle i|\}$ in  a fixed basis $I$ when measuring the state $|\psi\rangle$ is defined by,
\begin{equation}\label{randomness}
R_{I}(|\psi\rangle):=S(\Delta(\rho_{\psi})).
\end{equation}
For mixed state, the idea is generalized via purification procedure but it ultimately turns out to be the usual convex-roof extension, i.e.,
$$R_{I}(\rho)=\min_{\{p_e, |\psi_e\rangle\}}\sum_{e}p_e R_{I}(|\psi_e\rangle ),$$
where the minimization is over all pure state decompositions of the mixed state $\rho=\sum_{e}p_e|\psi_e\rangle \langle\psi_e|$ and $\sum_{e}p_e=1$.\\

Maximally coherent states \cite{preserving,MCS1} are the states which have the highest amount of resource. They attain maximum value for a coherence measure and it depends on the chosen measure. However, there is a set of states $\mathcal{M}$ which attain the maximum irrespective of the chosen measure,
\begin{equation}\label{maxst}
\mathcal{M}:=\{|\phi\rangle=\frac{1}{\sqrt{d}}\sum_{j=0}^{d-1} e^{\mathfrak{i}\theta_j}|j\rangle : \theta_j\in[0,2\pi] \}.
\end{equation}
This class is invariant under incoherent unitary operations. In fact this class is obtained from incoherent unitary operation $U=\sum_{j=0}^{d-1} e^{\mathfrak{i}\theta_j}|\alpha_j\rangle\langle j|$ ($\alpha_j$'s are simply reshuffling of $j$'s) on the state $|\phi_d\rangle:=\frac{1}{\sqrt{d}}\sum_{k=0}^{d-1} |k\rangle$.  \\

Suppose a quantum state is passing through a quantum channel. Cohering power \cite{coh pow1,coh pow2,coh pow3} refers to optimal creation of coherence by a quantum process and decoherening power refers to optimal destruction of coherence in a quantum process. Mathematically, cohering power of a quantum channel $\Lambda$ is defined as
\begin{equation}
\mathcal{C}_C(\Lambda):=\max_{\rho\in\mathcal{I}}\{C(\Lambda(\rho))-C(\rho)\}.
\end{equation}
Decohering power of a quantum channel $\Lambda$ is defined as
\begin{equation}\label{decohering power1}
\mathcal{D}_C(\Lambda):=\max_{\rho\in\mathcal{M}}\{C(\rho)-C(\Lambda(\rho))\},
\end{equation}
where $C$ stands for any coherence measure and $\mathcal{M}$ is the set of maximally coherent state. The definition (\ref{decohering power1}) reduces to
\begin{equation}\label{decohering power2}
\mathcal{D}_C(\Lambda)=1-\min_{\rho\in\mathcal{M}}C(\Lambda(\rho)),
\end{equation}
and it shows a kind of duality between these quantities. Results are available for specific qubit channels(such as unitary, depolarizing, bit-flip ) and few higher dimensional ones.

The concept of coherence distillation was  introduced by Yuan \textit{et al.}\cite{coh distillation} and Winter \textit{et al.}\cite{coh resource-win} in connection to resource theoretic formulation of coherence. Coherence distillation is
an asymptotic process by which maximally coherent state(considering $d=2$ in (\ref{maxst})) can be extracted from a non-maximally coherent state with incoherent operation. In other words, the distillable coherence of a state $\rho$(considering $d=2$) is the maximal rate at which a maximally coherent state $|\phi_2\rangle$ can be extracted from the given state. Let, $m$ copies of $|\phi_2\rangle$ can be extracted by incoherent operation from $n$ copies of the state $\rho$ in asymptotic sense with small positive $\epsilon$ error and $R=\frac{m}{n}$ is the rate of distillation. Then, distillable coherence of $\rho$ is defined as\cite{coh resource-win}
\begin{equation}
C_d(\rho):=\sup R,\quad \text{as} \quad n\rightarrow\infty, \epsilon\rightarrow 0.
\end{equation}
The distillation rate is equal to relative entropic coherence, giving it the operational significance, i.e., $C_d(\rho)=C_r(\rho)$. Again, for any pure state $|\psi\rangle$, distillation rate is approximately equal to intrinsic randomness \cite{intrand}, i.e., $R\approx R_{I}(|\psi\rangle)$.

\section{coherence fraction}
Let $\rho $ be any arbitrary state of dimension $d$. We define the coherence fraction of this state as
\begin{equation}\label{coh frac}
F_c(\rho):=\max_{|\phi\rangle\in\mathcal{M}}\langle\phi|\rho|\phi\rangle.
\end{equation}
where $\mathcal{M}$ is the set of maximally coherent states of same dimension. Coherence fraction of a state quantifies the maximal overlap of the state with maximally coherent state in a fixed basis. Thus, it quantifies the closeness of a state to maximally coherent state. It is always positive and normalized to $1$. Coherence fraction is maximum for maximally coherent states($\mathcal{M}$) and incoherent states have coherence fraction $\frac{1}{d}$. Since, coherence remains invariant under incoherent unitary operations, coherence fraction remains invariant under such unitary transformation. In fact, the definition of coherence fraction can be recast in another way as
\begin{equation}\label{coh frac1}
F_c(\rho)=\max_{U}\langle\phi_d|U\rho U^{\dagger}|\phi_d\rangle,
\end{equation}
where $U$ is any incoherent unitary operation and $|\phi_d\rangle=\frac{1}{\sqrt{d}}\sum_{k=1}^d |k\rangle$ is a particular representative from $\mathcal{M}$. This new definition resulted from the invariance of the set of maximally coherent states under the action of incoherent unitary.\\

We will now state a theorem which will connect coherence fraction with $l_1$-norm coherence. Here we again mention that throughout the paper we will consider coherence in the computational basis only.


\begin{theorem}\label{cf-cl1 theorem}
 Coherence fraction of an arbitrary state  $\rho $ of dimension $d$ is given by
\begin{equation}\label{cf-cl1 relation}
F_{c}(\rho)=\frac{1}{d}+\frac{1}{d}C_{l_1}(\rho).
\end{equation}
if and only if
\begin{equation}\label{state property}
\theta_{jp}+\theta_{pk}=2n\pi+\theta_{jk}
\end{equation}
  where $\theta_{jk}\in[0,2\pi]$ is the argument of $(j,k)$-th element of the density matrix $\rho$.
\end{theorem}
\textbf{\textit{Proof :}} Refer Appendix \ref{expression1}.\\

It is clear that for general mixed qudit states the right hand side of relation (\ref{cf-cl1 relation}) work only as an upper bound of coherence fraction. 
This theorem provide us a bound that is achieved by a large class of states, say, $\mathcal{A}$, with the property given in eq(\ref{state property}).
For example, any state with non-negative entries satisfies this bound.  Therefore coherence fraction of any state, satisfying the property(\ref{state property}), could be used to detect coherence: a state is coherent iff its coherence fraction is strictly greater than $\frac{1}{d}$ and incoherent iff the coherence fraction is exactly $\frac{1}{d}$. This theorem also provide us that the coherence fraction of pure states and X-states(whose density matrix contains only diagonal and main off diagonal elements) ( follows from the property given in eq(\ref{state property})).


Next, we consider a mixed qutrit state where the right hand side of relation (\ref{cf-cl1 relation}) works only as an upper bound. Let us  consider the following mixed qutrit state
$$\rho=(1-p)|\psi\rangle\langle\psi|+p|\phi\rangle\langle\phi|,$$
where $|\psi\rangle=\frac{1+\mathfrak{i}}{\sqrt{6}}|0\rangle+\frac{\sqrt{2}+\mathfrak{i}\sqrt{5}}{2\sqrt{3}}|1\rangle+\frac{1}{2\sqrt{3}}|2\rangle$ and $|\phi\rangle=\frac{\mathfrak{i}}{\sqrt{2}}|0\rangle+\frac{\sqrt{3}}{2\sqrt{2}}|1\rangle+\frac{1}{2\sqrt{2}}|2\rangle.$
We observe after performing $10^4$ simulations, that for $p=\frac{1}{2}$, $F_{c}(\rho)<\frac{1}{d}+\frac{1}{d}C_{l_1}(\rho)$.\\

Our next result will provide connection between coherence fraction and robustness of coherence.
\textbf{Corollary:} For any qubit state $\rho$,
\begin{equation}\label{cf-cR relation}
F_{c}(\rho)=\frac{1}{2}+\frac{1}{2}C_{\mathcal{R}}(\rho),
\end{equation}
where $C_{\mathcal{R}}(\rho)$ is the robustness of coherence (defined below).\\

Robustness of coherence (ROC) $C_{\mathcal{R}}(\rho)$ was introduced and studied in details in \cite{coh robustness}. For any state $\rho$, ROC is given by
$$C_{\mathcal{R}}(\rho)=\min_{\tau}\{s\ge0|\frac{\rho+s\tau}{1+s}=:\delta\in\mathcal{I}\},$$
where $\tau$ is any arbitrary state. ROC has a closed form  for qubit state and  $C_{\mathcal{R}}(\rho)=C_{l_1}(\rho)$. This clearly implies the above result. This measure of coherence is an important measure of coherence, having operational interpretation. It quantifies the advantage enabled by a quantum state, compared to any incoherent state, in phase discrimination task. Thus the  relation between coherence fraction and robustness of coherence provides the importance of coherence fraction in operational aspects also.\\

We are now going to state a result which will connect coherence fraction with coherence distillation which ultimately provide coherence fraction as an operationally meaningful quantity and relates it to the resource theory of coherence.

\begin{theorem}\label{cf-distillity relation}
An arbitrary  pure state of dimension $2$ is distillable iff its coherence fraction is strictly greater than $\frac{1}{2}$.
\end{theorem}

\textbf{\textit{Proof : }} Let us consider any pure state $\rho_{\psi}=|\psi\rangle\langle\psi|$ of dimension $2$. 
As distillable coherence is equivalent to the randomness for pure state, we can write $C_d(\rho_{\psi})=C_r(\rho_{\psi})=S(\Delta(\rho_{\psi}))=R_{\textit{I}}(\rho_{\psi})$ \cite{coh resource-win,coh distillation}.
Now, for any arbitrary qubit state $\rho$, randomness is given by\cite{coh distillation}
\begin{equation}\label{randomness and coh}
R_{\textit{I}}(\rho)=H\left(\frac{1+\sqrt{1-C_{l_1}^2(\rho)}}{2}\right),
\end{equation}
where $H(\cdot)$ denotes usual binary entropy. Therefore, using the result of Theorem \ref{cf-cl1 theorem}, we obtain
\begin{equation}\label{cf and distillation}
C_d(\rho_{\psi})=H\left(\frac{1+2\sqrt{F_{c}(\rho_{\psi})(1-F_{c}(\rho_{\psi}))}}{2}\right),
\end{equation}
The above relation indicates that whenever coherence fraction of any pure state $\rho_{\psi} $ is strictly greater than $\frac{1}{2}$, it is distillable and vice-versa.  Again, we have shown in Theorem \ref{cf-cl1 theorem}, that any qubit state is coherent iff its coherence fraction is strictly greater than $\frac{1}{2}$. Thus, we can say that every pure coherent state is distillable in qubit system. This result establishes coherence fraction as a faithful quantifier of the resource theory of coherence. This result can also be verified from  $\textit{Theorem 1}$ in \cite{coh transformation} and the result of Winter \textit{et al.}\cite{coh resource-win} for qubit system which states that there is no ``bound coherence''  from which no coherence could be distilled, i.e., in order to create it, nonzero coherence would be required.

\section{Optimal Coherence fraction of channel}
Having introduced the idea of coherence fraction of a quantum state, we are in a position to introduce another information theoretic quantifier. However, this quantifier represents properties of quantum channel.  Quantum noise is inevitable in any quantum process thus affecting the overall coherence of a quantum state. This new quantity will measure the ability of a channel in retaining coherence.
Let, $|\psi\rangle $ be a pure coherent state and it is affected by quantum noise which is modeled as quantum channel $\Lambda$. The noise transforms the initial pure state to a mixed state $\rho_{\psi,\Lambda}=\Lambda(\rho_{\psi})$ where $\rho_{\psi}=|\psi\rangle\langle\psi|$. Coherence fraction of the output state is given by
\begin{equation}\label{coh frac of output}
F_c(\rho_{\psi,\Lambda})=\max_{|\phi\rangle\in\mathcal{M}}\langle\phi|\rho_{\psi,\Lambda}|\phi\rangle.
\end{equation}
This quantity $F_c(\rho_{\psi,\Lambda})$ depends on the channel. It can increase or decrease according to the channel is cohering or decohering. It remains unaffected for coherence preserving channel and always attains the maximum value $1$. For completely decohering channel this quantity becomes constant $\frac{1}{d}$. If we vary the input state this quantity quantifies optimum power of the channel, affecting the resourcefulness in a quantum state in terms of coherence and it is the sole property of the channel. Thus, we define the optimal coherence fraction of a quantum channel $\Lambda$ as
\begin{equation}\label{opt coh frac}
\mathcal{F}_c(\Lambda)=\max_{|\psi\rangle}F_c(\rho_{\psi,\Lambda}).
\end{equation}
Identity channel, coherence preserving channel have optimal coherence fraction as 1 and decohering channel has optimal coherence fraction $\frac{1}{d}$. It is clear from the Eqs.(\ref{coh frac of output}) and (\ref{opt coh frac})
 \begin{equation}
\mathcal{F}_c(\Lambda)\ge F_c(\rho_{\psi,\Lambda}).
 \end{equation}
We will now try to obtain optimal coherence fraction of general qubit channel and some other well known qubit channels.

\subsection{Optimal coherence fraction for qubit channels}
Any qubit channel $\Lambda$ is a completely positive trace preserving (CPTP) map. Any completely positive, trace-preserving map on qubits can be represented by a matrix in the canonical basis of $\{I, \sigma_{1}, \sigma_{2},\sigma_{3}\}$ \cite{channel1} as
\begin{equation}\label{channel}
    \Lambda(\rho)=\frac{1}{2}[I+(\textbf{t}+T\textbf{v}).\pmb{\sigma}],~~~~~\text{where}~~ \rho=\frac{1}{2}[I+\bf{v}.\pmb{\sigma}]
\end{equation}
Here $\textbf{t}$ is the column vector  $(t_{1}, t_{2}, t_{3})^{t}$, $t_{k}\in \mathbb{R}$, $k=1,2,3 $ and $T=[T_{ij}]$ is a $3\times3$ matrix i.e., $\textbf{T}=\left(
    \begin{array}{cc}
    1 & \textbf{0} \\
    \textbf{t} & T \\
    \end{array}
            \right)$;
$\textbf{0}$ is a zero row vector of dimension 3. The map $\Lambda$ is called unital if and only if $\textbf{t} = \textbf{0}$.

\begin{theorem}\label{ocf-cl1 theorem}
The optimal coherence fraction of a qubit channel is given by
\begin{equation}\label{opt coh frac for qubit}
\mathcal{F}_{c}(\Lambda)=\frac{1}{2}+\frac{1}{2}\max_{|\phi\rangle\in \mathcal{M}}C_{l_1}(\rho_{\phi,\Lambda}),
\end{equation}
where $C_{l_1}(\rho_{\phi,\Lambda})$ is the $l_1$-norm coherence of the state $\rho_{\phi,\Lambda}$ and $\mathcal{M}$ denotes the set of maximally coherent states.
\end{theorem}

\textbf{\textit{Proof :}} Refer Appendix \ref{expression2}\\

We will now present a result which will connect the two related properties of channel -- one is optimal coherence fraction and another is decohering power of a channel.

\begin{theorem}\label{ocf-decoh pow th}
For any qubit channel $\Lambda$,
\begin{equation}\label{ocf-decoh pow relation}
2\leq 2\mathcal{F}_{c}(\Lambda)+D_{C_{l_1}}(\Lambda)\leq 3.
\end{equation}
\end{theorem}

\textbf{\textit{Proof :}} The proof follows from the corresponding definitions of the above two quantities. From the definitions, $$\mathcal{F}_{c}(\Lambda)=\frac{1}{2}+\frac{1}{2}\max_{|\phi\rangle\in \mathcal{M}}C_{l_1}(\rho_{\phi,\Lambda}),$$ and $$D_{C_{l_1}}(\Lambda)=1-\min_{|\phi\rangle\in \mathcal{M}}C_{l_1}(\rho_{\phi,\Lambda}).$$ It is obvious that  $$\max_{|\phi\rangle\in \mathcal{M}}C_{l_1}(\rho_{\phi,\Lambda})\ge \min_{|\phi\rangle\in \mathcal{M}}C_{l_1}(\rho_{\phi,\Lambda}).$$ Therefore,$$ 2\mathcal{F}_{c}(\Lambda)+D_{C_{l_1}}(\Lambda)=2+K,$$ where $K=\max_{|\phi\rangle\in \mathcal{M}}C_{l_1}(\rho_{\phi,\Lambda})-\min_{|\phi\rangle\in \mathcal{M}}C_{l_1}(\rho_{\phi,\Lambda})$. The definition of $K$ at once tells $0\le K\leq 1$. Lower bound is achieved by bit-flip, depolarizing and general amplitude damping channels.

\subsection{Optimal coherence fraction for some qubit channels}
We will now obtain the optimal coherence fraction for some qubit channels. We will discuss the complementary nature of optimal coherence fraction and decohereing power in each cases. We will start by the general unitary channel, which is fundamental to any quantum evolution.\\

\textbf{One-qubit unitary operation.}
The action of a general unitary qubit channel $\mathcal{U}$ on arbitrary qubit $\rho$ is given by $\mathcal{U}(\rho)=U\rho U^{\dagger}$ where $U=e^{\mathfrak{i}\frac{\phi}{2}\hat{\textbf{n}}.\pmb{\sigma}}$. Optimal coherence fraction of this channel is given by
\begin{equation}
\mathcal{F}_{c}(\mathcal{U})=\frac{1}{2}+\frac{1}{2}p_{+},
\end{equation}
and the decohering power of the channel is given by
\begin{equation}
D_{C_{l_1}}(\mathcal{U})=1-p_{-},
\end{equation}
where $p_{\pm}=\frac{X_1^2+X_2^2+Y_1^2+Y_2^2}{2}\pm\sqrt{\left(\frac{X_1^2+X_2^2+Y_1^2+Y_2^2}{2}\right)^2-(X_1^2X_2^2+Y_1^2Y_2^2)}$ and
$X_{1,2}=\cos{\phi}+2n_{1,2}^2\sin^2\frac{\phi}{2}$, $Y_{1,2}=2n_1n_2\sin^2\frac{\phi}{2}\pm n_3\sin{\phi}$. As $0\le p_{+}\le 1$, $\frac{1}{2}\le \mathcal{F}_{c}(\mathcal{U})\le 1$. $\mathcal{F}_{c}(\mathcal{U})$ attains its maximum value at $\hat{\textbf{n}}=(\pm1,0,0)$ or $(0,\pm1,0)$, $\phi=\pm\frac{\pi}{2}$ and $\hat{\textbf{n}}=(\pm\frac{1}{\sqrt{2}},0,\pm\frac{1}{\sqrt{2}})$ or $(0,\pm\frac{1}{\sqrt{2}},\pm\frac{1}{\sqrt{2}})$, $\phi=\pm\pi$ .
Thus   $\mathcal{F}_{c}(I\pm\mathfrak{i}\sigma_{1,2})=1$, $\mathcal{F}_{c}[\pm\mathfrak{i}(\sigma_{1,2}+\sigma_3)]=1$.\\

\textbf{Depolarizing channel.}
Evolution of a qubit $\rho$ under depolarizing channel ($\Lambda_{dep}$) is given by
\begin{equation}\label{dep channel}
\Lambda_{dep}(\rho)=(1-p)\rho+p\frac{I}{2}, \quad 0\le p\le 1.
\end{equation}
Simple calculations give optimal coherence fraction as
\begin{equation}
\mathcal{F}_{c}(\Lambda_{dep})=1-\frac{p}{2}.
\end{equation}
Therefore, $\frac{1}{2}\le \mathcal{F}_{c}(\Lambda_{dep})\le 1$. Corresponding to the completely depolarizing channel, the optimal coherence fraction attains the minimum and it shows the complete decohering nature of the channel. Again, we have $D_{C_{l_1}}(\Lambda_{dep})=p$. Thus, $2\mathcal{F}_{c}(\Lambda_{dep})+D_{C_{l_1}}(\Lambda_{dep})=2$. So, decohering channel attains the lower bound of the complementary relation in Theorem \ref{ocf-decoh pow th}. \\

\textbf{Bit-flip channel.}
Evolution of a qubit state $\rho$ under bit-flip channel ($\Lambda_{bf}$) is given by
\begin{equation}\label{bf channel}
\Lambda_{bf}(\rho)=(1-p)\rho+p\sigma_1\rho\sigma_1, \qquad 0\le p\le 1.
\end{equation}
Optimal coherence fraction of this channel always attains the maximum value, i.e., $ \mathcal{F}_{c}(\Lambda_{bf})=1, \:\forall p $.
Decohering power of the channel is $D_{C_{l_1}}(\Lambda_{bf})=1-|1-2p|$. Thus, $2\mathcal{F}_{c}(\Lambda_{bf})+D_{C_{l_1}}(\Lambda_{bf})=3-|1-2p|$. Hence, this is another extreme example of a channel for which the complementary relation attains both the lower and upper bounds.\\

\textbf{Generalized amplitude damping channel.}
The previous channels, we have discussed, were unital in nature. The channel we are considering here is non-unital in nature. Evolution of a qubit $\rho$ under general amplitude damping (GAD) channel ($\Lambda_{gad}$) is given by
\begin{equation}\label{GAD channel}
\Lambda_{gad}(\rho)=\sum_{i=1}^4E_i\rho E_i^{\dagger},
\end{equation}
where $\{E_i\}$ is the set of Kraus operators describing the GAD channel satisfying condition $\sum_i E_i^{\dagger}E_i=I$. $E_i$'s for GAD channel are given by follows

$E_1=\sqrt{\gamma}
\left( \begin{array}{cc}
  1 & 0 \\
  0 & \sqrt{1-p} \\
\end{array}
\right)
$,
$E_2=\sqrt{\gamma}
\left( \begin{array}{cc}
    0 & \sqrt{p} \\
    0 & 0 \\
\end{array}
\right)
$,
$E_3=\sqrt{1-\gamma}
\left( \begin{array}{cc}
      \sqrt{1-p} & 0 \\
      0 & 1 \\
\end{array}
\right)
$,
$E_4=\sqrt{1-\gamma}
\left( \begin{array}{cc}
        0 & 0 \\
        \sqrt{p} & 0 \\
\end{array}
\right)
$ \\
where $0\le p, \gamma \le 1$. For $\gamma=1$, the channel reduces to amplitude damping channel. We obtain the optimal coherence fraction of such channel as
\begin{equation}
\mathcal{F}_{c}(\Lambda_{gad})=\frac{1}{2}+\frac{1}{2}\sqrt{1-p}.
\end{equation}
Again, decohering power of such channel is $D_{C_{l_1}}(\Lambda_{gad})=1-\sqrt{1-p}$. In this case, we still get the lower bound corresponding to the complementary relation i.e., $2\mathcal{F}_{c}(\Lambda_{gad})+D_{C_{l_1}}(\Lambda_{gad})=2$.\\

\textbf{Self-complementary channel.} These are a class of special quantum channels those are equal to their complementary channels \cite{selfcomp}.  Hence the coherent information of such channel is zero, i.e., information does not leak out to the environment. Kraus operators corresponding to one-qubit self-complementary channel are:\\

$$E_1=
\left( \begin{array}{cc}
  1 & 0 \\
  0 & \frac{1}{\sqrt{2}}\sin \theta \\
\end{array}
\right),\,
E_2=
\left( \begin{array}{cc}
    0 & \frac{1}{\sqrt{2}}\sin \theta\\
    0 & e^{\mathfrak{i}\phi}\cos\theta \\
\end{array}
\right),
$$\\
and the free phases $\theta\in[0,\pi],\phi\in[0,2\pi]$. Optimal coherence fraction of such class of channels is $\mathcal{F}_{c}(\Lambda_{self})=\frac{1}{2}+\frac{1}{2\sqrt{2}}|\sin\theta|p_{max}$ and decohering power is $D_{C_{l_1}}(\Lambda_{self})=1-|\sin\theta|p_{min}$ where $p_{max}=\max\{|1+\cos{\theta}|,|1-\cos{\theta}|\}$, $p_{min}=\min\{|1+\cos{\theta}|,|1-\cos{\theta}|\}$. Hence the complementary relation shows $2\mathcal{F}_{c}(\Lambda_{self})+D_{C_{l_1}}(\Lambda_{self})=2+\frac{|\sin2\theta|}{2\sqrt{2}}$. So, it is an interesting class of channels for which the range $[2,2.35]$ of the expression (\ref{ocf-decoh pow relation}) is achieved.

\section{Extension to bipartite system}
We have introduced the concept of coherence fraction and optimal coherence fraction for single system. We can extend the scope to bipartite system too. The obvious necessary tools, i.e., the concept of coherence in bipartite system and the concept of bipartite maximally coherent state have already been developed \cite{cohmult1,cohmult2}. Based on these, we can define the coherence fraction of a bipartite state in similar way to that of single system. On the other hand, the idea of optimal coherence fraction can be trivially extended by considering the effects of individual local actions of channels.
\subsection{Coherence fraction}
Consider a bipartite state $\rho_{ab}$, shared between two parties, with local Hilbert space dimensions $d_1$ and $d_2$ respectively. The global coherence fraction can be defined as
\begin{equation}\label{coh frac for two qubit}
F_c(\rho_{ab})=\max_{|\phi\rangle\in\mathcal{M}_2}\langle\phi|\rho_{ab}|\phi\rangle,
\end{equation}
where $\mathcal{M}_2$ is  the set of  bipartite maximally coherent states , defined as \cite{MCS1}
$$\mathcal{M}_2:=\{|\phi\rangle=\frac{1}{\sqrt{d_1 d_2}}\sum_{k,l=0}^{d_1-1,d_2-1}e^{\mathfrak{i} \theta_{kl}}|kl\rangle: \theta_{kl}\in[0,2\pi]\}.$$
We can extend the result of Theorem \ref{cf-cl1 theorem}  in multiparty level too and connect $l_1$-norm coherence with the coherence fraction as
\begin{equation}\label{cf for bipartite pure state}
F_{c}(\rho_{{ab}})=\frac{1}{d_1 d_2}+\frac{1}{d_1 d_2}C_{l_1}(\rho_{{ab}}).
\end{equation}
where  $\rho_{ab}$ is any state which satisfies the property 
\begin{equation}\label{bipartite state property}
\theta_{ik}+\theta_{kj}=2n\pi +\theta_{ij}, 
\end{equation}
 $\theta_{ij}$ denotes the argument  of the $(i,j)$-th element of the density matrix $\rho_{ab}$.
 The proof is a trivial generalization of the previous result. This result gives a firm connection between coherence fraction and $l_1$-norm coherence in pure state  and $X$ state level. It at once follows that $ \frac{1}{d_1 d_2}\le F_{c}(\rho_{{ab}})\le 1$ and  $\rho_{{ab}}$ is coherent iff $F_{c}(\rho_{{ab}})> \frac{1}{d_1 d_2}$. The definition of coherence fraction in multiparty level gives us the access to compare the coherence fraction of a quantum state and its local parts just like quantum correlation measures. Here, we should careful about the fact that the concept of coherence fraction and entanglement fraction are qualitatively different things. The following example will show none could be derived from other.\\
Consider the two qubit state:
  $$\rho=p|++\rangle\langle++|+(1-p)|\Phi^+\rangle\langle\Phi^+|$$
 where $|+\rangle=\frac{|0\rangle+|1\rangle}{\sqrt{2}}$ and $|\Phi^+\rangle=\frac{|00\rangle+|11\rangle}{\sqrt{2}}, ~~ 0 \le p \le 1$.
 
 $C_{l_1}(\rho)=1+2p$ and $F_c(\rho)=\frac{1+p}{2}$.\\
 For $p=1$, this state is separable but maximally coherent.\\

Now, we are going to show a result which will connect local and global coherence fraction of this class of states.
\begin{theorem}
For a two qubit state satisfying property(\ref{bipartite state property}) 
\begin{equation}\label{relations between bipartite cf}
F_{c}(\rho_{{a}})+F_{c}(\rho_{{b}})\le 2F_{c}(\rho_{{ab}})+\frac{1}{2},
\end{equation}
where $\rho_{{a}}$, $\rho_{{b}}$ are the reduced local density matrices of the bipartite state  $\rho_{{ab}}$.
\end{theorem}

\textbf{\textit{Proof : }} We can write form Eq. (\ref{cf for bipartite pure state})
$$F_{c}(\rho_{{ab}})=\frac{1}{4}+\frac{1}{4}C_{l_1}(\rho_{{ab}}).$$ As  theorem 2 holds good for any qubit state, therefore,
$$F_{c}(\rho_{{x}})=\frac{1}{2}+\frac{1}{2}C_{l_1}(\rho_{{x}}),\qquad x\in\{a,b\}.$$
By definition of $l_1$-norm of coherence we have
$$C_{l_1}(\rho_{{a}})+C_{l_1}(\rho_{{b}})\le C_{l_1}(\rho_{{ab}}).$$
Substituting the expression of $l_1$-norm coherence in terms of coherence fraction in the last inequality, we obtain the result of this theorem.\\

Specifically, this theorem gives a bound of local coherence fraction in terms of global coherence fraction for two-qubit state satisfying property(\ref{bipartite state property}) 
 and thus it shows that coherence fraction is not exactly monotonic in usual sense. However, the global coherence put some restriction on the values of local coherence. We will now extend the notion of optimal coherence fraction to bipartite level.\\

\subsection{Optimal Coherence fraction}
\begin{figure}[h]
    \subfloat[]{\includegraphics[scale=0.5]{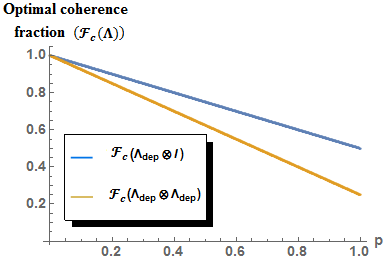}\label{ocf_dep}}\quad
    \subfloat[]{\includegraphics[scale=0.6]{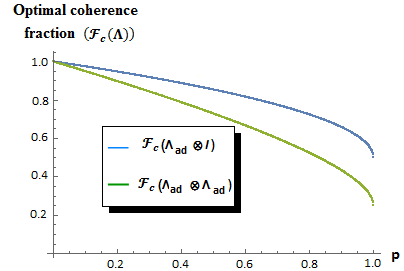}\label{ocf_ad}}\\
     \subfloat[]{\includegraphics[scale=0.6]{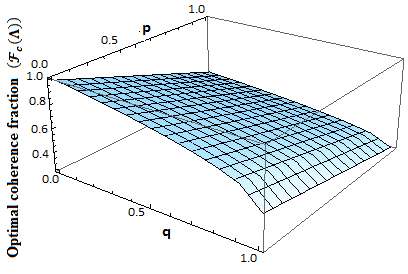}\label{ocf_depad}}\quad
     \subfloat[]{\includegraphics[scale=0.6]{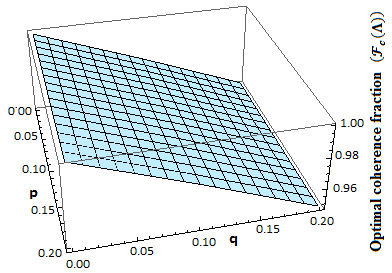}\label{ocf_bfad}}\\
     \caption{(Color online) Fig.(\ref{ocf_dep}) and (\ref{ocf_ad}) describe dynamics of optimal coherence fraction $\mathcal{F}_{c}$ with channel parameter $p$ for two qubit depolarizing and amplitude damping channel respectively. We have seen that in case of bipartite system, optimal coherence fraction due to one side channel application is greater than that due to both side channel applications.  Fig.(\ref{ocf_depad}) and (\ref{ocf_bfad}) describe dynamics of optimal coherence fraction due to two different channel application in two sides i.e., first one is for depolarizing($p$)-amplitude damping($q$) and  second one is for bit flip($p$)-amplitude damping($q$).}\label{fig44}
 \end{figure}

Consider a bipartite pure state $|\psi\rangle_{ab}$ and let the state be affected by the local noises $\Lambda_1$ and $\Lambda_2$ respectively on each local part.  This action results to a mixed state $\rho_{\psi_{ab},\Lambda_1\otimes \Lambda_2}=(\Lambda_1\otimes \Lambda_2)\rho_{\psi_{ab}}$.
Coherence fraction of the noisy state $\rho_{\psi_{ab},\Lambda_1\otimes \Lambda_2}$ is
$$F_c(\rho_{\psi_{ab},\Lambda_1\otimes \Lambda_2})=\max_{|\phi\rangle\in \mathcal{M}_2}\langle\phi|\rho_{\psi_{ab},\Lambda_1\otimes \Lambda_2}|\phi\rangle.$$
Then optimal coherence fraction corresponding to any two bipartite non-decohering quantum channels $\Lambda_1$, $\Lambda_2$ can be defined as
\begin{equation}
\mathcal{F}_c(\Lambda_1\otimes \Lambda_2)=\max_{|\psi\rangle_{ab}}F_c(\rho_{\psi_{ab},\Lambda_1\otimes \Lambda_2})
\end{equation}
\textbf{Numerical Results:} For two-qubit system, we have performed numerical simulation to find optimal coherence fraction by considering $10^3$ random values of channel parameters  for few well known channels like depolarizing, amplitude damping, bit flip.
We observe multiplicative nature of optimal coherence fraction under individual action of the local channels i.e.
$ \mathcal{F}_{c}(\Lambda_1\otimes \Lambda_2) =\mathcal{F}_{c}(\Lambda_1\otimes I) \mathcal{F}_{c}(I\otimes \Lambda_2).$
Similar to single qubit system, for bit-flip channel we get $\mathcal{F}_{c}(\Lambda_{bf}\otimes \Lambda_{bf})=\mathcal{F}_{c}(\Lambda_{bf}\otimes I)= \mathcal{F}_{c}(I\otimes \Lambda_{bf})=1 .$ Therefore we can say that the optimal coherence fraction is subadditive in nature.
Again, this quantity is symmetric, i.e., $\mathcal{F}_{c}(\Lambda_1\otimes \Lambda_2)=\mathcal{F}_{c}(\Lambda_2\otimes \Lambda_1)$, and also appending some extra local ancillary system does not affect the optimal coherence fraction of the channel i.e.
$\mathcal{F}_{c}(\Lambda\otimes I)=\mathcal{F}_{c}(\Lambda)$. The evolution of coherence fraction for different channel parameters and single side or both side use of channels are described in the Fig. \ref{fig44}. We particularly consider depolarizing and amplitude damping channel and observe the effect of channel parameters on optimal coherence fraction. It reveals that the application of the channel on both sides decreases the optimal coherence fraction.

\section{Conclusion}
In this work, we have defined two new quantities coherence fraction and optimal coherence fraction for both single and bipartite systems. For single system, we have shown that coherence fraction of a large class $\mathcal{A}$ of states, which includes all pure states and $X$ states  of dimension $d$, is a linear function of its  $l_1$-norm coherence. The same result holds good for large class of states in bipartite system too . 
The result works only as an upper bound for any mixed state other than that class of states. Even, we have found relations between coherence fraction with each of robustness of coherence and distillable coherence in qubit system. These relations provide the operational significance of quantum coherence: the advantage of coherent state in phase discrimination task and the distillability of any qubit coherent state. We have also shown a  type of complementary relationship between optimal coherence fraction and decohering power of a single qubit channel. Expressions of optimal coherence fraction  have been calculated for some well known qubit channels such as unitary channel, bit-flip channel, generalized amplitude damping channel, depolarizing channel and for a special type of channel that is self-complementary channel. We observe that  optimal coherence fraction is a linear function of channel parameter for depolarizing channel. On the other hand, for generalized amplitude damping channel, it is a nonlinear function of channel parameter. Coherence fraction in bipartite system obeys a monogamy type relationship with coherence fraction of local systems. In two-qubit system, numerical simulation for some qubit channels like bit-flip, amplitude damping and depolarizing channel shows that optimal coherence fraction is sub additive in nature. It is also symmetric and does not affected by adding the local ancilla.
The relation between coherence fraction and $l_1$ norm coherence has been studied in our work. Its relations with robustness of coherence and distillable coherence provide its operational importance. Such relationship between coherence fraction and other measures of coherence can be investigated further in future study. Specially there are scopes of studying the role of coherence fraction in describing and interpreting the scenario where coherence shows its advantages.  In literature\cite{duan}, we found examples where communication efficiency scales polynomially with the channel length which has a relation with entanglement fraction. We expect that in proper implementation, the coherence fraction will determine the amount of effectiveness in various experimental or communication schemes like entanglement fraction. For example, in the scheme of entanglement connection through swapping, described in\cite{duan}, they need some coherent conversion of atomic excitations into photons. The efficiency of the transfer could be related with our notion of coherence fraction.

\begin{acknowledgements}
The author S. Karmakar acknowledges the financial support from UGC, India. The author A.Sen acknowledges  NBHM, DAE India and the authors D. Sarkar and I. Chattopadhyay  acknowledges it as part of QUest initiatives by DST India.
\end{acknowledgements}

\begin{appendix}

\section{Proof of Theorem 1}\label{expression1}
\textit{
Coherence fraction of an arbitrary state  $\rho $ of dimension $d$ is given by
\begin{equation}\label{cf-cl1 relation appendix}
F_{c}(\rho)=\frac{1}{d}+\frac{1}{d}C_{l_1}(\rho).
\end{equation}
if and only if
\begin{equation}\label{state property appendix}
\theta_{jp}+\theta_{pk}=2n\pi+\theta_{jk}
\end{equation}
 where $\theta_{jk}\in[0,2\pi]$ is the argument  of $(j,k)$-th element of the density matrix $\rho$.
 }

\textbf{\textit{Proof :}} Any qudit state can be written as, $\rho =\sum_{j,k=0}^{d-1}\alpha_{jk}|j\rangle \langle k|$ where $\alpha_{jk}\in \mathbb{C}$ and $\sum_{j=0}^{d-1}|\alpha_{jj}|^2=1$.  Then it is straightforward to see that
$$F_{c}(\rho_{\psi})=\max_{}\frac{1}{d}+\frac{1}{d} \max_{\theta}\sum_{\substack{j,k=0 \\ j \ne k}}^{d-1} e^{-\mathfrak{i}(\theta_j-\theta_k)}\alpha_{jk},$$
where  $|\phi\rangle=\frac{1}{\sqrt{d}}\sum_{j=0}^{d-1} e^{\mathfrak{i}\theta_j}|j\rangle$ is the maximally coherent state of dimension $d$ and $\theta=(\theta_0,\theta_1,...,\theta_{d-1})$. 
Now, considering some simple mathematical inequalities we obtain
$$\sum_{\substack{j,k=0 \\ j \ne k}}^{d-1} e^{-\mathfrak{i}(\theta_j-\theta_k)}\alpha_{jk}\le |\sum_{\substack{j,k=0 \\ j \ne k}}^{d-1} e^{-\mathfrak{i}(\theta_j-\theta_k)}\alpha_{jk}|\le \sum_{\substack{j,k=0 \\ j \ne k}}^{d-1} |\alpha_{jk}|.$$
The inequalities also hold under optimization and thus
$$\max _{\theta}\sum_{\substack{j,k=0 \\ j \ne k}}^{d-1} e^{-\mathfrak{i}(\theta_j-\theta_k)}\alpha_{jk}\leq\sum_{\substack{j,k=0 \\ j \ne k}}^{d-1} |\alpha_{jk}|.$$ It is easy to calculate that this upper bound is achieved if and only if $ e^{\mathfrak{i}(\theta_j-\theta_k)}=\frac{\alpha_{jk}}{|\alpha_{jk}|}, j,k(j\ne k)=0,1,\ldots,d-1$ i.e., when $\theta_{jp}+\theta_{pk}=2n\pi+\theta_{jk}$ where $\alpha_{jk}=|\alpha_{jk}|e^{\mathfrak{i}\theta_{jk}}$  and $\theta_{jk}\in[0,2\pi]$.


Hence completes the proof.

\section{Proof of Theorem 3}\label{expression2}
\textbf{Theorem 3 :} \textit{The optimal coherence fraction of a qubit channel is given by
\begin{equation}
\mathcal{F}_{c}(\Lambda)=\frac{1}{2}+\frac{1}{2}\max_{|\phi\rangle\in \mathcal{M}}C_{l_1}(\rho_{\phi,\Lambda}),
\end{equation}
where $C_{l_1}(\rho_{\psi,\Lambda})$ is the $l_1$-norm coherence of the state $\rho_{\psi,\Lambda}$ and $\mathcal{M}$ denotes the set of maximally coherent states. }

\textit{Proof :} By Theorem {\ref{cf-cl1 theorem}}, for any qubit state  $\rho_{\psi,\Lambda}$, one can write $F_{c}(\rho_{\psi,\Lambda})=\frac{1}{2}+\frac{1}{2}C_{l_1}(\rho_{\psi,\Lambda})$ where $\Lambda$ is any qubit channel. Therefore
\begin{equation}\label{max}
\mathcal{F}_{c}(\Lambda)=\frac{1}{2}+\frac{1}{2}\max_{|\psi\rangle}C_{l_1}(\rho_{\psi,\Lambda}).
\end{equation}
In context of coherence generation and consumption, we can categorize a qubit channel as coherence increasing and non-increasing channel. If the channel is coherence increasing, it is obvious that $C_{l_1}(\rho_{\psi,\Lambda})$ attains its maximum for some maximally coherent state.

Now suppose the channel $\Lambda$ is coherence non-increasing channel. Therefore for any incoherent input state $\rho$, output state $\Lambda(\rho)$ must be incoherent which implies that $t_1=t_2=0$ and $T_{13}=T_{23}=0$ in Eq.(\ref{channel}). Then $C_{l_1}(\rho_{\psi,\Lambda})=\sqrt{{v'}_1^2+{v'}_2^2}$ where $\textbf{v}'=({v'}_1,{v'}_2,{v'}_3)^t=(\textbf{t}+T\textbf{v})$ in Eq.(\ref{channel}). Now, $C_{l_1}(\rho_{\psi,\Lambda})$ is maximum when input state $|\psi\rangle$ is a maximally coherent state. This completes the proof.

\end{appendix}
\end{document}